\documentstyle [multicol,prl,aps,epsfig] {revtex}
\tightenlines
\begin{document}
\preprint{UFIFT-HEP-05-08}
\draft
\date{May 5, 2005}
\title{Results of a Search for Cold Flows of Dark Matter Axions}

\author{L. Duffy, P. Sikivie, and D.B. Tanner}
\address{Physics Department, University of Florida, Gainesville, FL 32611}

\author{S. Asztalos, C. Hagmann, D. Kinion, L. J Rosenberg, K. van
Bibber, and D. Yu}
\address{Lawrence Livermore National Laboratory, Livermore, CA 94550}

\author{R.F. Bradley}
\address{National Radio Astronomy Observatory, Charlottesville, VA 22903}

\maketitle

\begin{abstract}
Theoretical arguments predict that the distribution of cold 
dark matter in spiral galaxies has peaks in velocity space 
associated with non-thermalized flows of dark matter particles.  
We searched for the corresponding peaks in the spectrum of 
microwave photons from axion to photon conversion in the 
cavity detector of dark matter axions.  We found none and 
place limits on the density of any local flow of axions as 
a function of the flow velocity dispersion over the axion 
mass range 1.98 to 2.17 $\mu$eV.

\end{abstract}


\begin{multicols}{2}

\tightenlines

\noindent{\bf I. Introduction}

\vskip0.2cm

The axion \cite{WW,PQ} is a light pseudo-scalar particle whose 
existence would explain why the strong interactions conserve 
the discrete symmetries P and CP in spite of the fact that the 
Standard Model as a whole violates those symmetries.  It is the
quasi-Nambu-Goldstone boson associated with the spontaneous 
breaking of a $U_{\rm PQ}(1)$ global symmetry. The axion mass is 
\begin{equation}
m_a \simeq 6 \cdot 10^{-6}~{\rm eV}~\left({10^{12}~GeV \over f_a}\right)
\label{mass}
\end{equation}
where $f_a$, called the axion decay constant, is of order the 
expectation value that breaks $U_{\rm PQ}(1)$.  All axion couplings 
are inversely proportional to $f_a$.  In particular, the coupling 
exploited in our experiment, of the axion to two photons, is given by
\begin{equation}
{\cal L}_{a\gamma\gamma} = g_\gamma~{\alpha \over \pi f_a}~a~
\vec{E}\cdot\vec{B}~~~\ .
\label{interaction}
\end{equation}
In Eq.~(\ref{interaction}), $a$, $\vec{E}$ and $\vec{B}$ are the axion 
and electromagnetic fields, and $g_\gamma$ is a model-dependent factor 
of order one.  In one class of representative benchmark models, called
KSVZ \cite{KSVZ}, $g_\gamma = - 0.97$.  In another class, called DFSZ 
\cite{DFSZ}, $g_\gamma = 0.36$.

In the $10^{-5}$ to $10^{-6}$ eV range, the axion is a well-established
cold dark matter (CDM) candidate \cite{vacmis,axrev}.  The present
velocity dispersion of cosmic axions is \cite{chang}
\begin{equation}
\delta v_a \sim 3 \cdot 10^{-17} c 
\left({10^{-5} {\rm eV} \over m_a}\right)^{5 \over 6}~~~\ .
\label{prim}
\end{equation}

In many discussions of cold dark matter detection it is assumed that the
distribution of CDM particles in galactic halos is isothermal \cite{WIMP}.
An isothermal distribution may be the outcome of an early epoch of
``violent relaxation" \cite{LB} of the galactic halo.  However, there are
excellent reasons to believe that a large fraction of the local density of
cold dark matter particles is in discrete flows with definite velocities
\cite{is}.  Indeed, because CDM has very low primordial velocity
dispersion and negligible interactions other than gravity, the particles
lie on a 3-dim. hypersurface in 6-dim.  phase-space.  Because the number
of particles involved is extremely large, this ``phase-space sheet" is
continuous; it cannot break.  Wherever a galaxy forms, the sheet winds up
in phase-space. This implies that the velocity spectrum of CDM particles
at any physical location is discrete, i.e., it is the sum of distinct
flows each with its own density and velocity.  The existence of discrete
flows is seen in numerical simulations of galactic halo formation when
care is taken to enhance the resolution in the relevant regions of
phase-space \cite{widr}.

Discrete flows (sometimes called `streams') are also produced when
satellites, such as the Sagittarius dwarf galaxy, are tidally disrupted
by the gravitational field of the Milky Way \cite{stiff,free}. N-body
simulations of galactic halo formation predict \cite{sim} that the Milky
Way halo contains hundreds of such satellites.

A consequence of the existence of discrete flows is the formation of
caustics \cite{crdm}.  If the Earth is sufficiently close to a caustic, 
the local density is dominated by a pair of flows and a high resolution 
search becomes a very powerful tool for discovering the axion signal.  A
recent paper \cite{MW} claims on observational grounds that we are in fact
very close to a caustic in the Milky Way halo and that the local dark
matter density is dominated by a pair of flows, with one member of the
pair contributing 75\% of the local density.  If this is the case, the
corresponding line is much narrower than a thermalized spectrum and hence
has much higher signal to noise in a high resolution search.

\vskip0.2cm

\noindent{\bf II. Experiment}

\vskip0.2cm

Dark matter axions can be detected by stimulating their conversion 
to microwave photons in a cavity permeated by a strong magnetic 
field \cite{sik,ADMX}.  The energy of the outgoing photon equals 
the total energy of the incoming axion:
\begin{equation}
h \nu = m_a~(c^2 + {1 \over 2} v^2)~~\ ,
\label{encon}
\end{equation}
where $v$ is the velocity of the converted axion in the frame 
of the detector.  Because galactic halo axions have velocities 
of order $10^{-3} c$, the photon frequencies are spread over 
$\Delta \nu_a = \nu_a/Q_a$ where $\nu_a \equiv m_a c^2 / h$ 
is the axion mass frequency and $Q_a$, called the axion quality 
factor, is of order $10^6$.

When $\nu_a$ is at the center of the bandwidth of the cavity, 
the power generated in the cavity from axion conversion is \cite{sik}
\begin{eqnarray}
P~&=&~0.2~10^{-26}~{\rm W}~\left({V \over 200~{\rm L}}\right)
\left({B_0 \over 7~{\rm T}}\right)^2 C
\left({g_\gamma \over 0.36}\right)^2 \cdot \nonumber\\ 
&\cdot&\left({\rho_a \over 0.5~10^{-24} {\rm g}/{\rm cm}^3}\right)
\left({m_a c^2 \over h~{\rm GHz}}\right)~\min(Q,Q_a)\ ,
\label{power}
\end{eqnarray}
where $V$ is the cavity volume, $B_0$ the magnetic field strength, 
$\rho_a$ the local density of halo axions, and $Q$ the loaded 
quality factor of the cavity.  $C$ is a form factor which is 
largest in the fundamental TM mode.  Because $\nu_a$ is unknown 
(except within three orders of magnitude), the detector is made 
tunable to explore a wide range of frequencies.  Tuning is achieved 
by displacing dielectric and metal rods inside the cavity.  We 
calculate the value of $C$ for each rod position by numerically 
simulating the cavity.  Over the range of frequencies reported upon 
here, 478 -- 525 MHz, $C$ varies from 0.42 to 0.38.  The cavity volume 
is 189 L.  Typical values for $B_0$ and $Q$ are 7.8 T and $7 \cdot 10^4$
respectively.  The output of the cavity is amplified, shifted down in
frequency by mixing with a local oscillator, digitized, and spectrum 
analyzed using the Fast Fourier Transform (FFT) algorithm at two 
different resolutions, which we call medium (MR) and high resolution 
(HR).  See ref. \cite{ADMX} for a detailed description of the apparatus.

In the MR channel, the spectrum taken at each cavity setting is the
average of $10^4$ individual spectra, each of which is the FFT of the
cavity output voltage measured over an 8 ms time interval, for a total
measurement integration time of 80 s \cite{ADMX}.  The resulting
resolution (125 Hz) is well matched to the expected width $\Delta \nu_a =
\nu_a/Q_a$ (approx. 500 Hz for $\nu_a$ = 500 MHz) of the axion signal.  
The MR search is not predicated on any assumption about the velocity
distribution of axions in the Galactic halo, other than the very
conservative assumption that their velocity dispersion is not larger than
about 300 km/s.  

\begin{figure} 
\centerline {\epsfysize=2.3in
\epsfbox{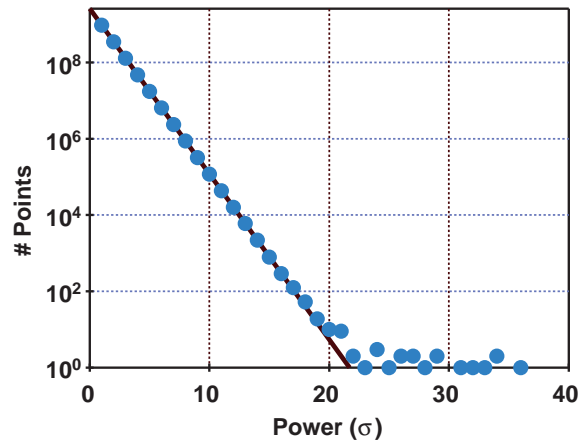} } 
\vspace{0.5cm} 
\caption{Power distribution for a large sample of single-bin data.} 
\label{stat1}
\end{figure}

In the HR channel we increase the resolution to 0.019 Hz by taking, at
each cavity setting, the FFT of a single 52 s long sequence of cavity
output measurements.  The sampling rate is set by the cavity bandwidth
$\nu/Q \simeq 6$ kHz (for $\nu$ = 500 MHz).  The FFTs involve therefore
approximately one million data points each.  A discrete CDM flow produces
a peak in the spectrum of microwave photons from axion conversion in our
detector.  Each peak's has a daily modulation due to the Earth's rotation
and an annual modulation due to its orbital motion \cite{FS}.  The latter
causes a peak frequency to shift by an amount of order 100 Hz or less in
the course of the year, whereas the daily shift is of order 1 Hz or less
(for $\nu_a = 500$ MHz). During the time (52s) spent taking data at each
cavity tune, a peak's frequency shifts by an amount which is at most $2
\cdot 10^{-3}$ Hz, due to the Earth's rotation.  The resulting broadening
of peaks is less than the present HR resolution (0.02 Hz).

We now describe the method by which we identify HR candidate peaks  
and discriminate spurious peaks (environmental and statistical) 
from potentially real axion signals.  At the highest resolution 
(the ``1-bin search"), the power $p_1$ in each frequency bin is 
the sum of independent sine and cosine components.  The {\it observed} 
probability distribution of $p_1$ is shown in Fig.~\ref{stat1}.  For 
low values of $p_1$, the distribution is of the form
\begin{equation}
\frac{dP}{dp_{1}}=\frac{1}{\sigma}
        \exp\left(-\frac{p_{1}}{\sigma}\right)~~~ \ .
\label{exp}
\end{equation}
This exponential behaviour is characteristic of the thermal and electronic 
noise from the cavity/amplifier chain.  $\sigma$ is the average noise 
power in the 1-bin search. Hence, $\sigma$ is proportional to the
total noise temperature $T_{\rm n} = T_{\rm ph} + T_{\rm el}$, where 
$T_{\rm ph}~(\simeq$ 1.7 K) is the physical temperature of the cavity 
and $T_{\rm el}~(\simeq$ 2.0 K) is the total electronic noise temperature 
of the receiver chain.  We verified this relationship experimentally 
by heating the cavity and showing that $\sigma$ is a linear function 
of $T_{\rm ph}$.  Indeed, that measurement constitutes our calibration 
of the power emitted by the cavity.

\begin{figure}
\centerline {\epsfysize=3.7in \epsfbox{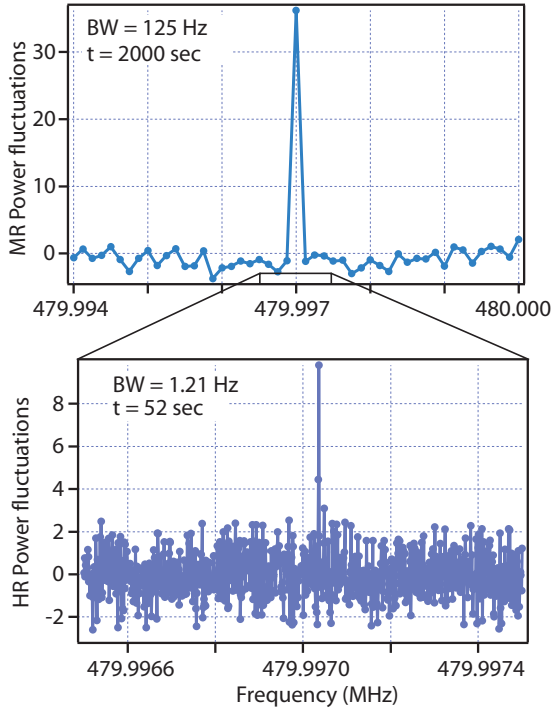} }
\vspace{0.5cm}
\caption{An environmental peak as it appears in the MR search (top) 
and the 64-bin HR search.  The unit for the vertical axis is the
rms power fluctuation in each case.}
\label{crossc}
\end{figure}

Thermal and electronic noise is only one part of the background in the
search for an axion signal.  The other part is a set of narrow lines from
the R.F. environment of the experiment.  These are signals from nearby
emitters ({\it e.g.} computer clocks) which leak into the cavity by a
variety of means.  We refer to such narrow lines as `environmental
peaks'.  The difference between the observed $p_1$-distribution and
Eq.~(\ref{exp}) for $p_1 \gtrsim 20 \sigma$ (see Fig.~\ref{stat1}) is 
due to environmental peaks.

We performed a comparison between the signal power observed in 
the HR and MR channels using an environmental peak at 480 MHz; 
see Fig. \ref{crossc}. The HR and MR signals were determined to 
have power $1.8\cdot10^{-22}$ W and $1.7\cdot10^{-22}$ W in that 
peak respectively.  The difference between the two measurements 
is consistent with the size of the noise fluctuations in the HR
channel.  Note that this MR spectrum was acquired over a much 
longer measurement integration time than the HR spectrum (2000 s 
vs. 52 s).

In addition to the 1-bin search, we conduct $n$-bin searches with 
$n = 2,~4,~8,~64$, 512 and 4096, to search for dark matter axion 
flows of correspondingly larger velocity dispersion.  An $n$-bin 
is the sum of $n$ adjacent 1-bins.  Each $n$-bin is made to overlap 
with half of the previous $n$-bin to avoid missing a peak which would 
otherwise spread over adjacent $n$-bins.  

Before candidate peaks are selected, each HR spectrum is corrected for the
combined passband-filter response of the receiver chain and for the
frequency-dependent response of the coupling between the cavity and the
first cryogenic amplifier.  These corrections are similar to those for the
MR channel, and have been described in detail elsewhere \cite{ADMX}.  
For each corrected HR spectrum, the $p_1$-distribution is fitted to
Eq.~(\ref{exp}) for $0 < p_1 \lesssim 1\sigma$, to determine the value of
$\sigma$.  The 2-bin spectrum is obtained by combining adjacent bins in
the 1-bin spectrum, the 4-bin spectrum is likewise obtained from the 2-bin
spectrum, and so on.  For each $n$, a power threshold was chosen such 
that any peak with power larger than that threshold will be considered 
a candidate axion signal.  The thresholds were set as low as possible
consistent with the requirement that the number of candidates remain
manageable.  The thresholds are 20, 25, 30, 40, 120, 650 and $4500 \sigma$
for $n$ = 1, 2, 4, 8, 64, 512 and 4096 respectively.

The set of candidate peaks that passed these thresholds were investigated 
in subsequent runs, using the same methods as we apply to the MR channel 
\cite{ADMX}.  All our candidates were shown to be environmental peaks. 

We now place a limit on the density of any persistent axion dark 
matter flow as a function of its velocity dispersion, over the 
frequency range covered (478 -- 525 MHz).  The corresponding axion 
mass range is 1.997 to 2.171 $\mu$eV.  The $n$-bin search places a 
limit on an axion flow of velocity disperion equal to or less than
\begin{equation}
\delta v_n =~12~{{\rm m} \over {\rm s}}~n~
\left({300~{\rm km/s} \over v}\right)~~~\ ,
\label{veldis}
\end{equation}
where $v$ is the flow velocity in the laboratory reference frame.
Eq.~(\ref{veldis}) was obtained by varying Eq.~(\ref{encon}) and 
using $\delta \nu_n = n$ (0.019 Hz) for the bin width, and 
$\nu = 500$ MHz for the frequency.  

The cavity was operated at near-critical coupling.  Hence, half of 
the power of Eq.~(\ref{power}) is measured by the receiver chain (the 
other half being lost in the cavity walls), provided (1) the axion 
mass frequency $\nu_a$ falls at the center of the cavity bandwidth, 
and (2) the frequency of the axion signal peak is at the center of a
single bin.  

If the axion mass frequency $\nu_a$ differs from the cavity resonant 
frequency $\nu_0$, the observed power is reduced by the Lorentzian 
factor
\begin{equation}
h(\nu_a) = {1 \over 1 + 4Q^2({\nu_a \over \nu_0} - 1)^2}~~~~\ .
\label{Lorentz}
\end{equation}
To set conservative limits, we use for each cavity setting the smallest
value of the Lorentzian factor over the bandwidth covered by that setting.

If the frequency $\nu$ of an axion signal peak does not fall at the center
of a bin, the signal power is spread over several bins.  It can be shown
that the minimum fraction of the power in a narrow line that ends up 
in a single $n$-bin is 40.5\% for $n=1$, 81\% for $n=2$, 87\% for $n=4$,
and 93\% for $n=8$.  For $n=64$, 512 and 4096, practically all the power 
ends up in a single bin regardless of where $\nu$ falls relative to the
bin boundaries.  To set conservative limits, we use the above minimum
fractions of signal power that fall in a single $n$-bin.

\begin{figure}
\centerline {\epsfysize=2.6in \epsfbox{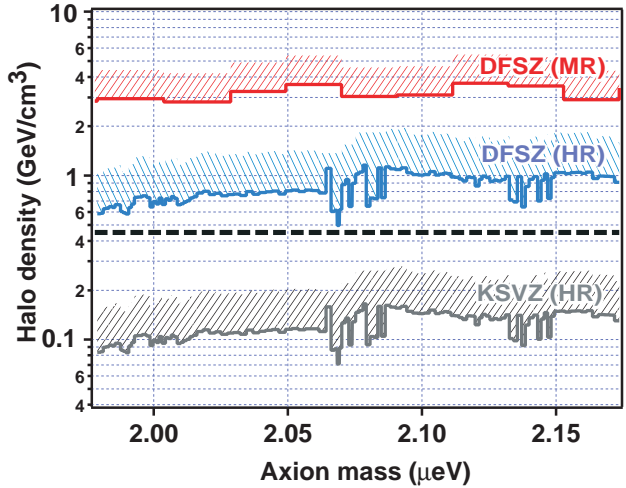} }
\vspace{1.0cm}
\caption{97.7\% confidence level limits from the HR 2-bin search
on the density of any local axion dark matter flow as a function
of axion mass, for the DFSZ and KSVZ $a\gamma\gamma$ coupling 
strengths. Also shown is the previous ADMX limit using the MR 
channel, for DFSZ coupling.  The HR limits assume that the flow 
velocity dispersion is less than $\delta v_2$ given in
Eq.~(\ref{veldis}).  The dotted line is the local dark 
matter density estimate given in ref. [20].}
\label{lim2}
\end{figure}

For the 64, 512 and 4096-bin searches, a background subtraction was
performed.  For those searches, the signal power levels such that the sum
of signal plus background has 97.7\% probability to exceed the thresholds
stated earlier, beyond which a peak is considered a candidate axion
signal, are respectively 71, 182 and 531 $\sigma$.  The latter are
therefore the 97.7\% confidence level upper limits on an axion 
signal obtained from the 64, 512 and 4096-bin searches.  The limits 
from the lower $n$ searches are not appreciably improved by performing 
a background subtraction.  Those limits equal the thresholds (20, 25, 
30, and 40 $\sigma$ for $n$ = 1, 2, 4, and 8) beyond which a peak is
considered an axion signal candidate.

\vskip0.2cm

\noindent{\bf III. Discussion}

\vskip0.2cm

The limits on the density of a cold flow of axion dark matter derived 
from the 2-bin search are shown in Fig.~\ref{lim2}.  The limits derived 
from the other $n$-bin searches differ from the $n=2$ limits only by 
constant (frequency independent) factors.  These factors are respectively 
1.60, 1.00, 1.12, 1.39, 2.53, 5.9 and 17.2 for $n$ = 1, 2, 4, 8, 64, 512 
and 4096.  The limits from the $n \geq 4$ searches are less severe than 
the $n=2$ limit, but they are valid for larger velocity dispersions [see 
Eq.~(\ref{veldis})].  The $n=1$ limit is both less severe and less general
than the $n=2$ limit.

Fig.~\ref{lim2} shows the potential of the High Resolution analysis 
for finding or excluding cosmic axions when a significant fraction 
of the local density is contained in one or few lines resulting from 
the incomplete thermalization of infalling dark matter.   Over the
measured mass range, flows are excluded for densities of order 
1 GeV/cm$^3$ even for DFSZ axions, which has the smallest 
axion-photon coupling ($g_\gamma$ = 0.36), at 97.7\% c.l. 
This marks an improvement in sensitivity of a factor 3 over 
our previous Medium Resolution analysis.  

This work is supported in part by U.S. Department of Energy under 
Contract W-7405-ENG-48 at Lawrence Livermore National Laboratory, 
and under grant DE-FG02-97ER41029 at the University of Florida.

\end{multicols}{2}

\end{document}